%% file: paper5.tex
\documentclass{mecbic}
\providecommand{\anno}{2011}
\providecommand{\firstpage}{25}
\providecommand{\lastpage}{39}
\usepackage{breakurl, color}             
\usepackage{verbatim}
\usepackage{subfigure}
\usepackage{epsfig}
\usepackage{amsfonts}
\usepackage{amssymb}
\usepackage{amsmath}
\usepackage{listings}
\input{macros}

\newtheorem{theorem}{Theorem}[section]

\newtheorem{example}[theorem]{Example}

\title{A Spatial Calculus of Wrapped Compartments\thanks{This research is funded by the BioBITs Project (\emph{Converging
Technologies} 2007, area: Biotechnology-ICT), Regione Piemonte.}}
\author{Livio Bioglio$^1$, Cristina Calcagno$^{1,2}$, Mario Coppo$^1$, Ferruccio Damiani$^1$,\\ Eva Sciacca$^1$, Salvatore Spinella$^1$, Angelo Troina$^1$
\institute{$^1$Dipartimento di Informatica, Universit\`a di Torino}
\institute{$^2$Dipartimento di Biologia Vegetale, Universit\`a di Torino}
}
\def\titlerunning{A Spatial Calculus of Wrapped Compartments}
\def\authorrunning{L.~Bioglio, C.~Calcagno, M.~Coppo, F.~Damiani,\\\ E.~Sciacca, S.~Spinella and A.~Troina
}
\begin{document}
\maketitle
\pagestyle{plain}
\pagenumbering{arabic}
\setcounter{page}{25}

\begin{abstract}
The Calculus of Wrapped Compartments (CWC) is a recently proposed modelling language for the representation and simulation of biological systems behaviour. Although CWC has no explicit structure modelling a spatial geometry, its compartment labelling feature can be exploited to model various examples of spatial interactions in a natural way. However, specifying large networks of compartments may require a long modelling phase. In this work we present a surface language for CWC that provides basic constructs for modelling spatial interactions. These constructs can be compiled away to obtain a standard CWC model, thus exploiting the existing CWC simulation tool. A case study concerning the modelling of Arbuscular Mychorrizal fungi growth is discussed.
\end{abstract}

\section{Introduction}
\label{sect_intro}
\input{intro}

\section{The Calculus of Wrapped Compartments}
\label{sect_cwc}

\input{cwc_calculus}

\section{A Surface Language}
\label{sect_sl}
\input{surf_language_ma}


\section{Case Study: A Growth Model for AM Fungi}
\label{sect_ex}
\input{example}

\section{Conclusions and Related Works}
\label{sect_conc}
\input{conclu}

\bibliographystyle{eptcs}
\bibliography{fmb,am}

\begin{appendix}
\section*{Appendix: Implementation of the Surface Language}
\label{implementation}
\input{implementation}

\end{appendix}

\end{document}

%% file: macros.tex
\newif\ifmc
\mcfalse 

\newcommand{\lb}[1]
{\ifmc{\color{cyan}{#1}}\else{#1}\fi}

\newcommand{\qqop}[1]{\mathrel{\makebox[2em]{$#1$}}}

\newcommand{\agr}{\quad\big|\quad}

\newcommand{\AT}{\mathcal{A}}
\newcommand{\LT}{\mathcal{L}}

\newcommand{\LeftPat}{p}

\newcommand{\RightPat}{o}

\newcommand{\TOP}{\top}

\newcommand{\ov}[1]{\overline{#1}}

\newcommand{\srew}[1]{\stackrel{#1}{\longmapsto}}

\renewcommand{\wr}[3]{( #1 \into #2)^{#3}}

\makeatletter
\def\mapstofill@{%
\arrowfill@{\mapstochar\relbar}\relbar\rightarrow}
\newcommand*\xmapsto[2][]{%
\ext@arrow 0000\mapstofill@{#1}{#2}}
\makeatother

\newcommand{\srewrites}[1]{\stackrel{#1}{\longmapsto}}

\newcommand{\into}{\ensuremath{\,\rfloor}\,}

\newcommand{\conc}{\;\,}
\newcommand{\emptyseq}{\bullet}

\newcommand{\red}{\longmapsto}

%% file: intro.tex
Several complex biological phenomena include aspects in which space plays an essential role, key examples are the growth of tissues and organisms, embryogenesis and morphogenesis processes or cell proliferation.
This has encouraged, in recent years, the development of formal models for the description of biological systems in which spatial properties can be taken into account~\cite{CG10,barbuti2011SCLS,MV10}, as required by the emerging field of {\it spatial systems biology}~\cite{SMG11} which aims at integrating the analysis of biological systems with spatial properties

The  Calculus of Wrapped Compartments (CWC)~\cite{preQAPL2010,HCWC_mecbic10,CDDGGT_TCSB11} is a calculus for the description of biochemical systems which is based on the notion of a compartment which represents, in some sense, the abstraction of a region with specific properties (characterized by a {\it label}, a {\it wrap} and a {\it content}). Biochemical transformations are described via a set of stochastic reduction rules which characterize the behaviour of the represented system.

In a recent work~\cite{spatial_COMPMOD11} we have have shown how CWC can be used to model spatial properties of biological systems. The idea is to exploit the notion of compartment to represent spatial regions (with a fixed, two-dimensional topology) in which the labels plays a key role in defining the spatial properties. In this framework, the movement and growth of system elements are described, via specific rules (involving adjacent compartments) and the functionalities of biological components are affected by the spatial constraints given by the sector in which they interact with other elements.
CWC allows to model several spatial interactions in a very natural way. However, when the complexity of simulation scenarios increases, the specification of large networks of compartments each one having its own peculiar behaviour and initial state may require a long and error prone modelling phase.

In this paper we introduce a surface language for CWC that defines a framework in which the notion of space is included as an essential component of the system. The space is structured as a square grid, whose dimension must be declared as part of the system specification.  The surface language provides basic constructs for modelling spatial interactions on the grid. These constructs can be compiled away to obtain a standard CWC model, thus exploiting the existing CWC simulation tool.

A similar approach can be found in \cite{MV10} where the topological structure of the components is expressed via explicit links which require ad-hoc rules to represent movements of biological entities and a logic-oriented language to flexibly specify complex simulation
scenarios is provided.
{
In order to deal with larger biological systems, we are planning to extend the CWC to the spatial domain incrementally.
At this early stage we neglected to consider problems related to the
increase of the spatial rules with the increasing dimension of the grid\footnote{note that in a 2D model the space-related rules grow according to the square of the grid dimension}.
A partial solution to this problem is the use of appropriate data structures to represent entities scattered on a grid. A further step in this direction should be that of allowing the definition of different topological representations  for spatial distributions of the biological entities, like in~\cite{spatial_COMPMOD11}. This requires, obviously, that also the surface language be enriched with primitives suitable to express different spatial topology and related concepts (like the notion of proximity of locations and that of movement in space). The right spatial topology could also help to minimize the number of spatial rules needed for modeling phenomena.
A more ambitious goal will be that of providing a basis
for computational geometry to our simulator, in order to identify
spatially significant events for the simulation. This will requires however a much bigger implementation effort.
}

\paragraph{Organisation of the Paper} Section~\ref{sect_cwc} recalls the CWC framework. Section~\ref{sect_sl} presents the surface language needed to describe spatial terms and rules. Section~\ref{sect_ex} presents a case study concerning some spatial aspects in the modelling of Arbuscular Mychorrizal fungi. Section~\ref{sect_conc} concludes the paper by briefly discussing related work and possible directions for further work. The Appendix presents the software module implementing the surface language.

%% file: cwc_calculus.tex
The Calculus of Wrapped Compartments (CWC)
(see~\cite{preQAPL2010,CDDGGT_TCSB11,HCWC_mecbic10}) is based on a nested structure of ambients delimited by membranes with specific proprieties. Biological entities
like cells, bacteria and their interactions can be easily described in CWC.

\subsection{Term Syntax}
\label{CWC_formalism - syntax}

Let $\AT$ be a set of  {\it atomic elements} ({\it atoms} for
short), ranged over by $a$, $b$, ..., and  $\LT$ a set of {\it compartment types} represented as {\it labels} ranged over by $\ell,\ell',\ell_1,\ldots$
A {\it term} of CWC is a multiset $\ov{t}$ of  {\it simple terms} where a simple term  is either an atom $a$ or a compartment $(\overline{a}\into
\overline{t'})^\ell$ consisting of a {\it wrap} (represented by the multiset of atoms $\overline{a}$), a {\it content} (represented by the term
$\overline{t'}$) and a {\it type} (represented by the label $\ell$).

\begin{figure*}
\centering
\subfigure[] {
\includegraphics[height=28mm]{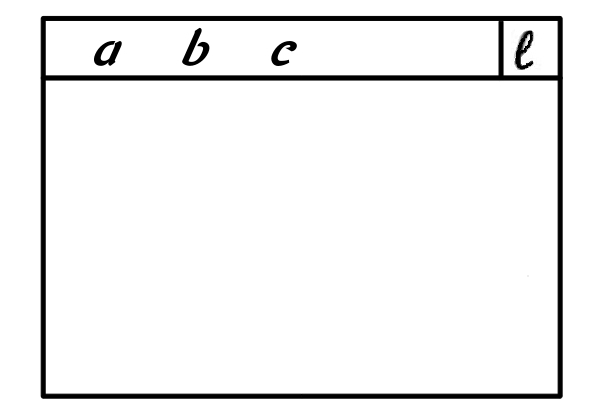}
}
\centering
\subfigure[] {
\includegraphics[height=28mm]{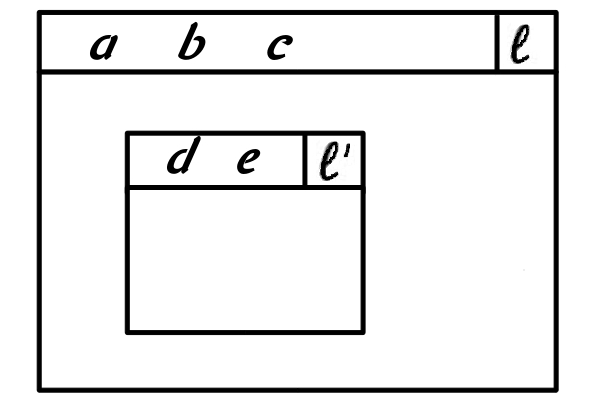}
}
\subfigure[] {
\includegraphics[height=28mm]{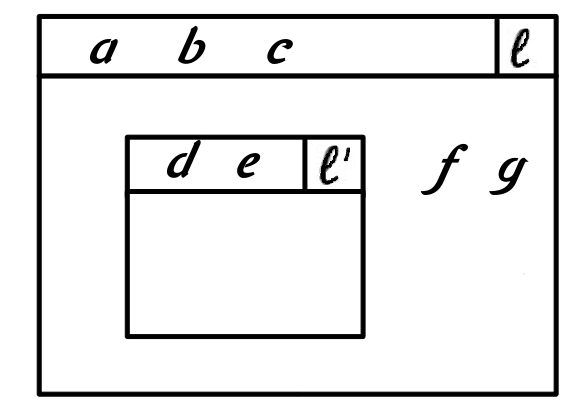}
}

\caption{\textbf{(a)} represents $(a \conc b \conc c \into \emptyseq)^\ell$; \textbf{(b)} represents $(a \conc b \conc c  \into (d \conc e \into
\emptyseq)^{\ell'})^\ell$; \textbf{(c)} represents $(a \conc b \conc c \into (d \conc e \into \emptyseq)^{\ell'} \conc f \conc g)^\ell$}
\label{fig:example CWM}
\end{figure*}

As usual, the notation $n*t$ denotes $n$ occurrences of the simple term $t$. We denote an empty term with $\emptyseq$. An example of CWC term is $2\!*\!a \conc b \conc (c \conc d \into e \conc
f)^\ell$ representing a multiset (multisets are denoted by listing the elements separated by a space) consisting of two occurrences of $a$, one occurrence
of $b$ (e.g. three molecules) and an $\ell$-type compartment $(c \conc d \into e \conc f)^\ell$ which, in turn, consists of a wrap (a membrane) with two
atoms $c$ and $d$ (e.g. two proteins) on its surface, and containing the atoms $e$ (e.g. a molecule) and $f$ (e.g. a DNA strand). See
Figure~\ref{fig:example CWM} for some other examples with a simple graphical representation.

\medskip

\subsection{Rewriting Rules}
\label{CWC_formalism - rewriting}

System transformations are defined by rewriting rules, defined by resorting to CWC terms that may contain variables.
We call {\it pattern} the l.h.s. component $\ov{\LeftPat}$ of a rewrite rule and {\it open term} the r.h.s. component $ \ov{\RightPat}$ of a rewrite rule, defined as multiset of {\it simple patterns} $p$ and {\it simple open terms} $o$ given by the following syntax:
$$
\begin{array}{lcl}
   \LeftPat  & \;\qqop{::=}\; & a \agr (\overline{a} \conc x \into \overline{\LeftPat}\conc X)^\ell \\
   \RightPat & \;\qqop{::=}\; & a \agr (\ov{q} \into \ov{o})^\ell   \agr X\\
   q         & \;\qqop{::=}\; & a \agr x
\end{array}
$$
where $\overline{a}$ is a multiset of atoms,  $\overline{\LeftPat}$ is a pattern (a, possibly empty, multiset of simple patterns), $x$ is a {\it wrap
variable} (can be instantiated by a multiset of atoms), $X$ is a {\it content variable}
 (can be instantiated by a CWC term), $\ov{q}$ is a multiset of atoms and wrap variables and $\overline{\RightPat}$ is an open term (a, possibly empty, multiset of simple open terms). Patterns are intended to match, via substitution of variables with ground terms (containing no variables),
with compartments occurring as subterms of the term representing the whole system. Note that we force {\it exactly} one variable to occur in each
compartment content and wrap of our patterns and simple patterns.
 This prevents ambiguities in the instantiations needed to match a given compartment.\footnote{
 The linearity condition, in biological terms, corresponds to excluding that a transformation can depend on the presence of two (or more)
 identical (and generic) components in different compartments (see also~\cite{OP11}).}

A {\it rewrite rule} is a triple $(\ell, \ov{\LeftPat},\ov{\RightPat})$, denoted by $\ell:  \overline{\LeftPat}  \srew{}    \ov{\RightPat}$, where $\ov{\LeftPat}$ and $\ov{\RightPat}$ are such that 
the variables occurring in $\ov{\RightPat}$ are a subset of the variables occurring in $\ov{\LeftPat}$.
 %
The application of a rule $\ell:  \ov{\LeftPat} \red \ov{\RightPat}$ to a term~$\ov{t}$ is performed in the following way: 1) Find in $\ov{t}$ (if it
exists) a compartment of type $\ell$ with content $u$ and a substitution $\sigma$ of variables by ground terms such that $u = \sigma( \ov{\LeftPat} \conc
X)$\footnote{The implicit (distinguished) variable $X$ matches with all the remaining part of the compartment content.} and 2) Replace in $\ov{t}$ the
subterm $u$ with $\sigma(\ov{\RightPat}\conc X)$. We write  $\ov{t} \red \ov{t'}$ if $\ov{t'}$ is obtained  by applying a rewrite rule to  $\ov{t}$.
The rewrite rule  $\ell:  \ov{\LeftPat} \red \ov{\RightPat}$ can be applied to any compartment of type $\ell$ with $\overline{\LeftPat}$ in its content (that will be rewritten with $\overline{\RightPat}$).

For instance, the rewrite rule
  $\ell: a \conc b \red c$
means that in all compartments of type $\ell$ an occurrence of $a \conc b$  can be replaced by $c$

While the rule does not change the label $\ell$ of the compartment where the rule is applied, it may change all the labels of the compartments occurring in its content. For instance, the rewrite rule  $\ell: (a \conc x \into X)^{\ell_1} \red (a \conc x \into X)^{\ell_2}$ means that, if contained in a compartment of type $\ell$, all compartments of type~$\ell_1$ and containing an $a$ in their wrap can change their type to $\ell_2$.

For uniformity reasons we assume that the whole system is always represented by a term consisting of a single compartment with distinguished label $\top$ and empty wrap, i.e., any system is represented by a term of the shape $\wr{\emptyseq}{\ov{t}}{\top}$, which will be also written as $\ov{t}$, for simplicity.

\subsection{Stochastic Simulation}\label{SECT:STO_SEM}

A stochastic simulation model for biological systems can be defined
along the lines of the one
presented by Gillespie in \cite{G77}, which is, {\it de facto}, the standard way to model quantitative aspects of biological systems. The basic idea of
Gillespie's algorithm is that a rate function is associated with each considered chemical reaction which is used as the parameter of an exponential distribution modelling the probability that he reaction takes place. In the standard approach this reaction rate is obtained by multiplying the kinetic
constant of the reaction by the number of possible combinations of reactants that may occur in the region in which the reaction takes place, thus modelling the law of mass action. For rules defining spatial movement the kinetic constant can be interpreted as the speed of the movement. In \cite{preQAPL2010}, the reaction rate is defined in a more general way by associating to each reduction rule a function which can also define rates based on different principles as, for instance, the Michaelis-Menten nonlinear kinetics.

For simplicity, in this paper, we will follow the standard approach in defining reaction rates. Each reduction rule is then enriched by the kinetic constant $k$ of the reaction that it represents (notation $\ell:  \ov{\LeftPat} \srewrites{k} \ov{\RightPat} $).
For instance in evaluating the application rate of the stochastic rewrite rule $R=
\ell: a \conc b  \srewrites{k} c $ (written in the simplified form) to the term $\ov{t}=a\conc a \conc b \conc b$ in a compartment of type $\ell$ we must consider the number of the possible combinations of reactants of
the form $a\conc b$ in $\ov{t}$. Since each occurrence of $a$ can react with each occurrence of $b$, this number is 4. So the application rate of $R$ is
$k\cdot 4$.

\subsection{The CWC simulator}
\label{sec:simulator}

The CWC simulator~\cite{HCWC_SIM} is a tool under development at the Computer Science Department of the Turin University, based on Gillespie's direct method
algorithm~\cite{G77}. It treats CWC models with different rating semantics (law of mass action, Michaelis-Menten kinetics, Hill equation) and
it can run independent stochastic simulations over CWC models, featuring deep parallel optimizations for multi-core platforms on the top of
FastFlow~\cite{fastflow:web}. It also performs online analysis by a modular statistical framework.

%% file: surf_language_ma.tex
In this section we embed CWC into a surface language able to express, in a synthetic form, both spatial (in a two-dimensional grid) and biochemical CWC transformations. The semantics of a surface language model is defined by translation into a standard CWC model.

We distinguish between two kind of compartments: 
 \begin{enumerate}
\item {\it Standard} compartments (corresponding to the usual CWC compartments), used to represent entities (like bacteria or cells) that can move through space.
\item {\it Spatial} compartments, used to represent portions of space. Each spatial compartment defines a location in a two dimensional grid
through a special atom, called {\it coordinate}, that occurs on its wrap. A coordinate is denoted by \texttt{row.column}, where \texttt{row} and
\texttt{column} are intergers. Spatial compartments have distinguished labels, called {\it spatial labels}, that can be used to provide a specific
characterisation of a portion of space.
\end{enumerate}
For simplicity we assume that the wraps of each spatial compartment contains only the coordinate. Therefore, spatial compartment differentiations can be
expressed only in terms of labels.\footnote{Allowing the wrap of spatial compartments to contain other atoms, thus providing an additional mean to express
spatial compartment differentiations, should not pose particular technical problems (extend the rules of the surface language to deal with a general wrap
content also for spatial compartments should be straightforward).}

For example, the spatial compartment $( \texttt{1.2}  \into 2*b)^{\textit{soil}}$ represents the cell of the grid located in the first row and the second
column, and has type \textit{soil}, the spatial compartment $( \texttt{2.3}  \into 3*b \conc c)^{\textit{water}}$ represents a \textit{water}-type spatial
compartment in position \texttt{2.3}. In our grid we assume that molecules can float only through neighbor cells: all the rules of interaction between
spatial compartments must obviously contain the indexes of their location. For example, the rule $\TOP : ( \texttt{1.2} \conc x \into a\conc
X)^{\textit{water}} ( \texttt{2.2} \conc y \into Y )^{\textit{soil}} \srewrites{k} ( \texttt{1.2} \conc x \into X)^{\textit{water}}  ( \texttt{2.2}\conc y
\into a\conc Y)^{\textit{soil}}$ moves the molecule $a$ from the \textit{water} compartment in position \texttt{1.2} to the \textit{soil} compartment in
position \texttt{2.2} with a rate $k$ representing in this case, the speed of the movement of $a$ in downwards direction from a cell of
\textit{water}-type to a cell of \textit{soil}-type.

Let \texttt{R} and \texttt{C} denote the dimensions of our
\texttt{R} $\times$ \texttt{C} grid defined by \texttt{R} rows and
\texttt{C} columns. To increase the expressivity of the language we define a few structures to denote portions (i.e. sets of cells) of
the grid. With $\Theta$ we denote a  set of coordinates of
the grid and we use the notion \texttt{r.c} $\in \Theta$
when the coordinate~\texttt{r.c} is contained in the set
$\Theta$.  We define rectangles by
\texttt{rect}[\texttt{r.c},\texttt{r'.c'}] where
\texttt{r.c},\texttt{r'.c'} represent the edges of the rectangle. We
project rows and columns of our grid with the constructions
\texttt{row}[$i$] and \texttt{col}[$j$] respectively.

\begin{example}
The set $\Theta=\{\mathtt{6.6}\}\cup \mathtt{rect}[\mathtt{1.1},\mathtt{3.2}]
\cup \mathtt{col}[5]$ represents the set of coordinates
$$\Theta = \{ \mathtt{6.6}\} \cup \{ \mathtt{1.1}, \mathtt{2.1}, \mathtt{3.1}, \mathtt{1.2}, \mathtt{2.2} , \mathtt{3.2}\} \cup \{ \mathtt{i.5} \;| \; \forall i \in [1,\mathtt{R}]\}.$$
\end{example}

Note that \texttt{row}[$i$] is just a shorthand for
\texttt{rect}[\texttt{i.1},\texttt{i.\texttt{C}}]. Similarly for
columns.\\
  We use \texttt{[*]} as shorthand to indicate the whole grid
(i.e. \texttt{rect[1.1,R.C]}).

We also define four {\it direction} operators,
\texttt{N,~W,~S,~E} that applied to a range of cells shift them,
respectively, up, left, down and right.
 For instance \texttt{E(1.1)} = \texttt{1.2}. In the intuitive way, we also define the four diagonal movements (namely, \texttt{NW},~\texttt{SW},~\texttt{NE},~\texttt{SE}). With $\Delta$ we denote a set of directions and we use the special symbol $\diamond$ to denote the set containing all eight possible directions.

We convene that when a coordinate, for effect of a shit, goes out of
the range of the grid the corresponding point is eliminated from the
set.

\subsection{Surface Terms}

We define the initial state of the system under analysis as a set of
compartments modelling the two-dimensional grid containing the
biological entities of interest.

Let $\Theta$ denote a set of coordinates and $\ell_s$ a spatial label. We use the notation:
    $$\Theta, \ell_s  \boxplus \; \ov{t} $$
to define a set of cells of the grid. Namely $\Theta, \ell_s  \boxplus \; \ov{t}$ denotes the top level CWC term:
 $$\wr{\emptyseq}{   \wr{ \mathtt{r_1.c_1}}{ \ov{t}   }{\ell_s} \conc \ldots \conc \wr{ \mathtt{r_n.c_n}}{ \ov{t}  }{\ell_s}   }{\top}  $$
 where  $\mathtt{r_i.c_j}$ range over all elements of $\Theta$.

A spatial CWC term is thus defined by the set of grid cells covering
the entire grid.

\begin{figure}
\centering
\subfigure[Initial state] {
\includegraphics[width=.45\textwidth]{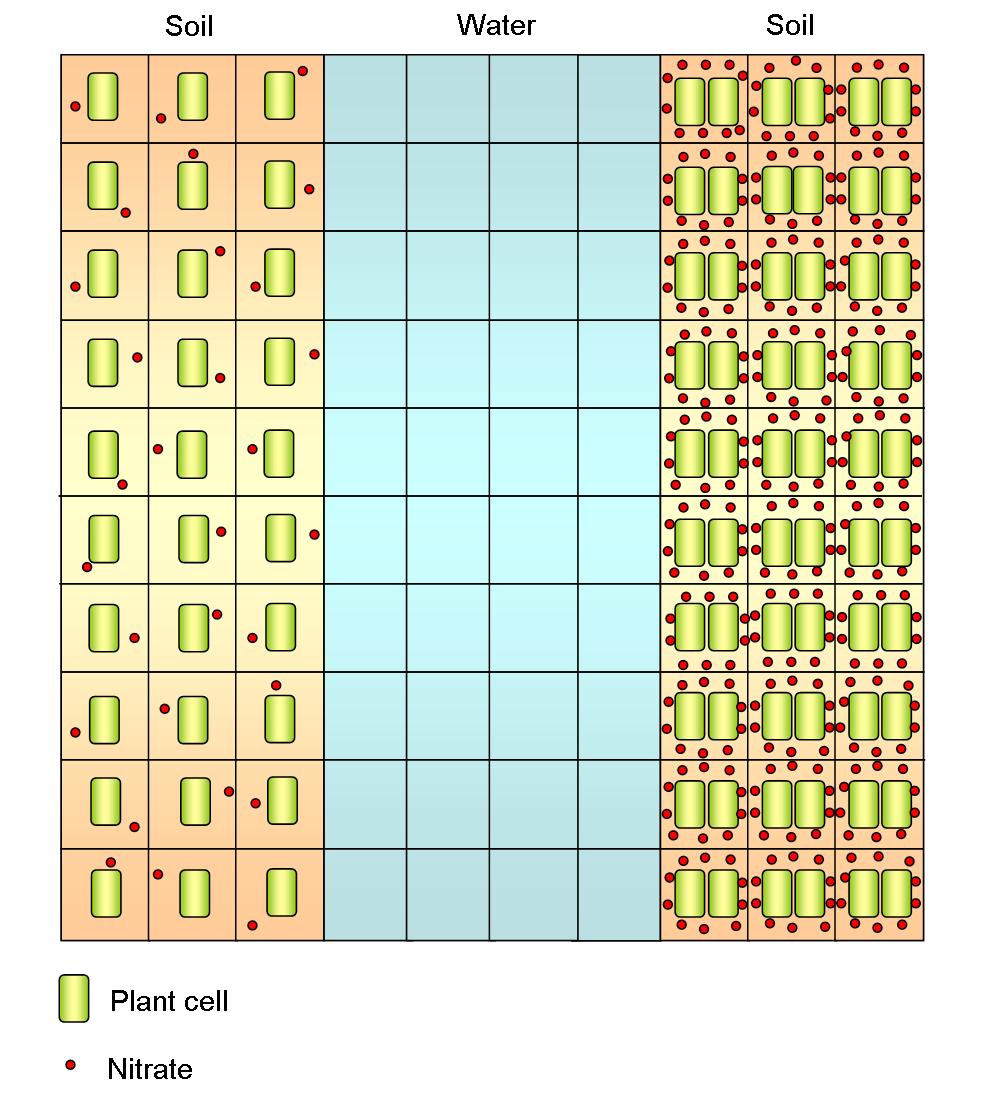}
\label{ex_grid_1}
}
\centering
\subfigure[Spatial events] {
\includegraphics[width=.45\textwidth]{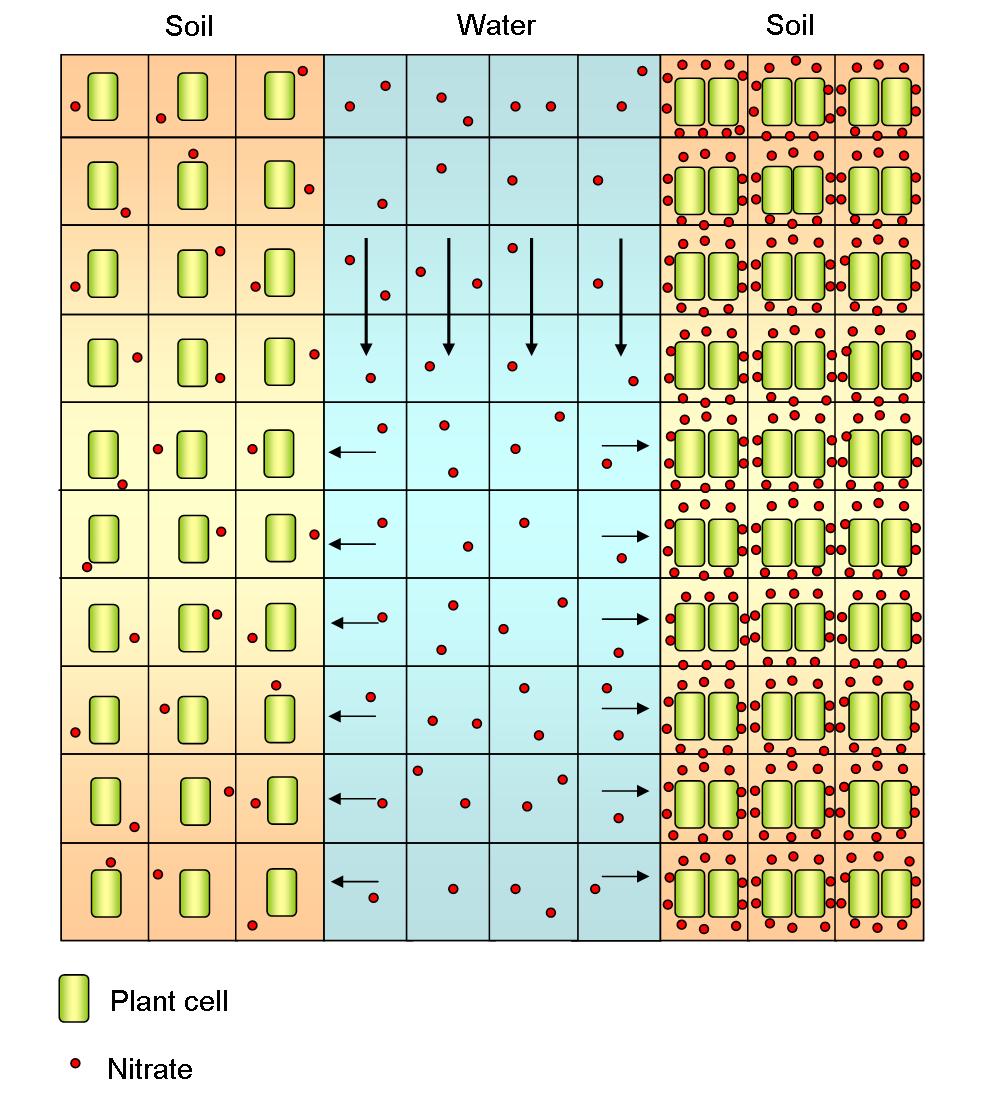}
\label{ex_grid_2}
}
\caption{Graphical representations of the grid described in the Example $3.2$.}\label{ex_grid}
\end{figure}

\begin{example}\label{EX_grid}
The CWC term obtained by the three grid cell constructions:\\

$
\mathtt{rect[1.1,10.3]}, soil \boxplus \; nitr \conc (receptors \into cytoplasm\conc nucleus)^{PlantCell}$\\
$\mathtt{rect[1.4,10.7]}, water \boxplus \; \emptyseq$ \\
$\mathtt{rect[1.8,10.10]}, soil \boxplus \; 10*nitr \conc 2 * (receptors \into cytoplasm \conc nucleus)^{PlantCell}$\\
\noindent builds a $10\times 10$ grid composed by two portions of soil (the right-most one reacher of nitrates and plant cells) divided by a river of water (see Figure \ref{ex_grid_1}).
\end{example}

\subsection{Surface Rewrite Rules}

We consider rules for modelling three kind of events.

\smallskip
\noindent
{\it Non-Spatial Events:} are described by standard CWC rules, i.e. by rules of the shape:
 $$\ell:  \ov\LeftPat \srewrites{k} \ov\RightPat $$
\noindent Non-spatial rules can be applied to any compartment of type $\ell$ occurring in any
portion of the grid and do not depend on a particular location.

\begin{example}
A plant cell might perform its activity in any location of the grid. The following rules, describing some usual activities within a cell, might happen in any spatial compartment containing the plant cells under considerations:\\
$PlantCell: nucleus  \srewrites{k_1} nucleus \conc  mRNA$\\
$PlantCell: mRNA \conc cytoplasm  \srewrites{k_2} mRNA \conc  cytoplasm \conc protein$.
\end{example}

\smallskip \noindent
{\it Spatial Events:} are described by rules that can be applied to specific spatial compartments. These rules allow to change the spatial label of the considered compartment. Spatial events are described by rules of the following shape:
$$
\Theta \triangleright \ell_s : \overline{\LeftPat} \srewrites{k}  \ell_s': \ov\RightPat
$$
\noindent Spatial rules can be applied only
within the spatial compartments with coordinates contained in the set~$\Theta$ and with the spatial label $\ell_s$. The application of the rule may also change the label of the spatial compartments $\ell_s$ to $\ell_s'$.
This rule is translated into the CWC set of rules:\\
$\TOP : (  \mathtt{r_i.c_i} \conc x \into
\overline{\LeftPat}\conc X)^{\ell_s} \srewrites{k} (  \mathtt{r_i.c_i}\conc x
\into \ov{\RightPat}\conc X )^{\ell_s'} \quad \forall \mathtt{r_i.c_i} \in
\Theta.$

Note that spatial rules are analogous to non spatial ones. The only difference is the explicit indication of the set $\Theta$ which allows to write a single rule instead of a set of rule (one for each element of $\Theta$).

\begin{example}
If we suppose that the river of water in the middle of the grid defined in Example~\ref{EX_grid} has a downward streaming, we might consider the initial part of the river (framed by the first row {\it \texttt{rect[1.4,1.7]}})
 to be a source of nitrates (as they are coming from a region which is not modelled in the actual considered grid). The spatial rule:\\
$
\mathtt{rect[1.4,1.7]}  \triangleright water: \emptyseq \srewrites{k_3} water: nitr
$\\
\noindent models the arrival of nitrates at the first modeled portion of the river (in this case it does not change the label of the spatial compartment involved by the rule).
\end{example}

 \noindent
{\it Spatial Movement Events:} are described by rules considering the content of two adjacent spatial compartments and are described by rules of the following shape:
$$
\Theta \triangleleft \Delta \triangleright \ell_{s_1}, \ell_{s_2}: \ov{p_1} ,\ov{p_2} \srewrites{k} \ell_{s_1}', \ell_{s_2}': \ov{o_1} ,\ov{o_2}
$$
\noindent This rule changes the
content of two adjacent (according to the possible directions contained in $\Delta$) spatial
compartments and thus allows to define the movement of objects. The
pattern matching is performed by checking the content of a spatial
compartment of type $\ell_{s_1}$ located in a portion of the
grid defined by $\Theta$ and the content of the adjacent spatial
compartment of type $\ell_{s_2}$. Such a rule could also change the
labels of the spatial compartments.
This rule is translated into the CWC set of rules:
$$\TOP : ( \mathtt{r_i.c_i}\conc x \into
\overline{\LeftPat_1} \conc X)^ {\ell_{s_1}}  ( dir(\mathtt{r_i.c_i}) \conc y \into
 \overline{\LeftPat_2} \conc Y) ^{\ell_{s_2}}
\srewrites{k} ( \mathtt{r_i.c_i} \conc x \into  \ov{o_1}\conc X)
^{\ell_{s_1}'}  ( dir(\mathtt{r_i.c_i}) \conc y \into   \ov{o_2}\conc Y )^{\ell_{s_2}'}$$
for all $\mathtt{r_i.c_i} \in \Theta$ and for all $dir \in \Delta$.

\begin{example}
We assume that the flux of the river moves the nitrates in the water according to a downward direction in our grid and with a constant speed in any portion of the river with the following rule:\\
$
\mathtt{rect[1.4,9.7]} \triangleleft \{\mathtt{S}\} \triangleright water, water: nitr,\emptyseq \srewrites{k_4} water, water: \emptyseq, nitr
$\\
when nitrates reach the down-most row in our grid they just disappear (non moving event):\\

$
\mathtt{rect[10.4,10.7]} \triangleright water: nitr \srewrites{k_4} water: \emptyseq
$.\\
Moreover, nitrates streaming in the river may be absorbed by the soil on the riverside with the rule:\\
$
\mathtt{rect[1.4,10.7]} \triangleleft \{\mathtt{W}, \mathtt{E}\} \triangleright water, soil: nitr,\emptyseq \srewrites{k_5} water, soil: \emptyseq, nitr
$.\\
A graphical representation of these events is shown in Figure \ref{ex_grid_2}. Other rules can be defined to move the nitrates within the soil etc.
\end{example}

%% file: example.tex
In this section we illustrate a case study concerning the modelling of Arbuscular Mychorrizal fungi growth.

\subsection{Biological Model}

The arbuscular mycorrhizal (AM) symbiosis is an example of association with high compatibility  formed between fungi belonging to the Glomeromycota phylum and the roots of most land plants\cite{harrison2005signaling}. AM fungi are obligate symbionts, in the absence of a host plant, spores of AM fungi germinate and produce a limited amount of mycelium.
The recognition between the two symbionts is driven by the perception of diffusible signals and once reached the root surface the AM fungus enters in the root, overcomes the epidermal layer and it grows inter-and intracellularly all along the root in order to spread fungal structures.
Once inside the inner layers of the cortical cells the differentiation of specialized, highly branched intracellular hyphae called arbuscules occur. Arbuscules are considered the major site for nutrients exchange between the two organisms. The fungus supply the host with essential nutrients such as phosphate, nitrate and other minerals from the soil. In return, AM fungi receive carbohydrates derived from photosynthesis in the host.

\begin{figure}
\center
\includegraphics[width=.8\textwidth]{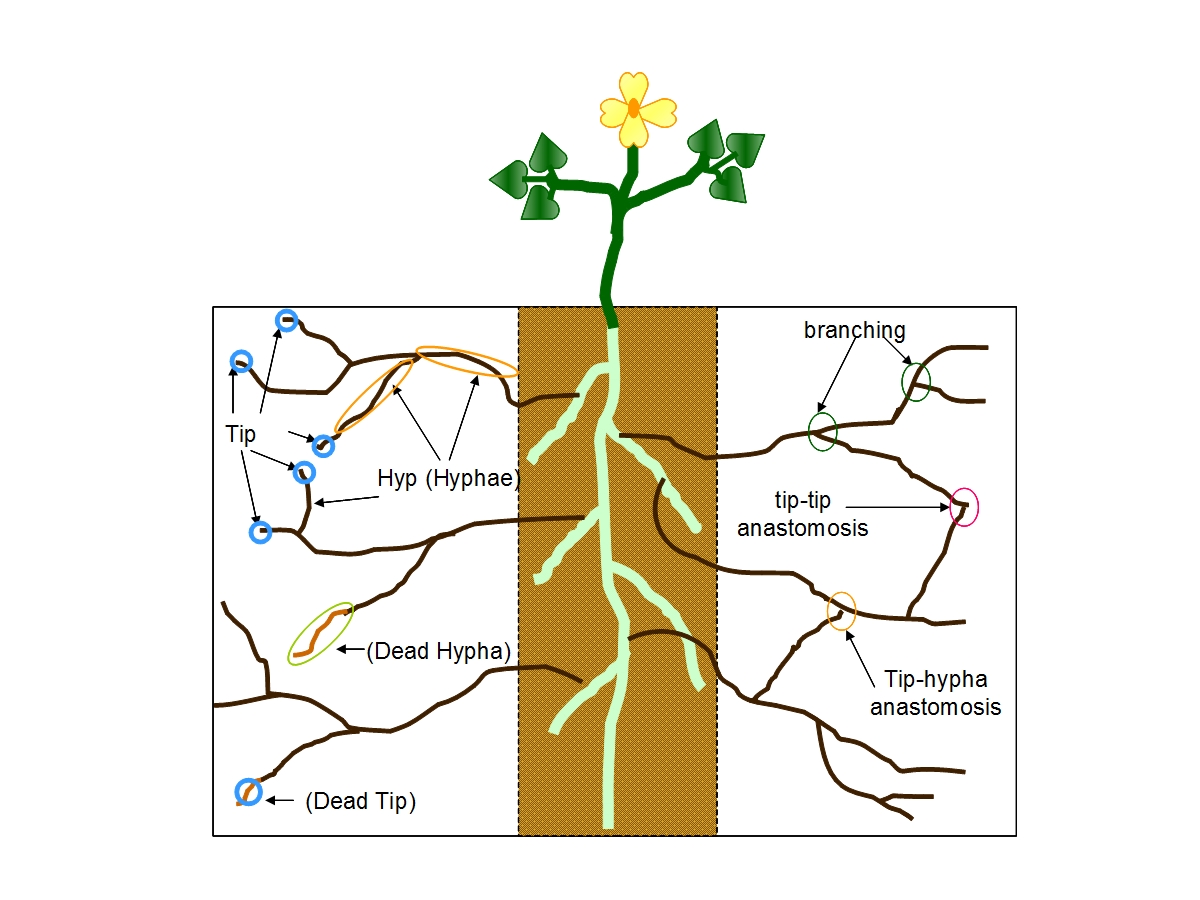}
\caption{Extraradical mycelia of an arbuscular mycorrhizal fungus.}\label{fig:ERM}
\end{figure}

Simultaneously to intraradical colonization, the fungus develops an extensive network of hyphae which explores and exploits soil microhabitats for nutrient acquisition. AM fungi have different hyphal growth patterns, anastomosis and branching frequencies which result in the occupation of different niche in the soil and probably reflect a functional diversity \cite{maherali2007influence} (see Figure \ref{fig:ERM}).
The mycelial network that develops outside the roots is considered as the most functionally diverse component of this symbiosis. Extraradical mycelia (ERM) not only provide extensive pathways for nutrient fluxes through the soil, but also have strong influences upon biogeochemical cycling and agro-ecosystem functioning \cite{purin2008parasitism}. The mechanisms by which fungal networks extend and function remain poorly characterized. The functioning of ERM presumably relies on the existence of a complex regulation of fungal gene expression with regard to nutrient sensing and acquisition.
The fungal life cycle is then completed by the formation, from the external mycelium, of a new generation of spores able to survive under unfavourable conditions.

Investigations on carbon (C) metabolism in AM fungi have proved useful to offer some explanation for their obligate biotrophism. As mentioned above, an AM fungus relies almost entirely on the host plant for its carbon supply. Intraradical fungal structures (presumably the arbuscules) are known to take up photosynthetically fixed plant C as hexoses. Unfortunately, no fungal hexose transporter-coding gene has been characterized yet in AM fungi.

In order to quantify the contribution of arbuscular mycorrhizal (AM) fungi to plant nutrition, the development and extent of the external fungal mycelium and its nutrient uptake capacity are of particular importance. Shnepf and collegues \cite{schnepf2008growth} developed and analysed a model of the extraradical growth of AM fungi associated with plant roots considering the growth of fungal hyphae from a
cylindrical root in radial polar coordinates.

Measurements of fungal growth can only be made in the presence of plant. Due to this practical difficulty experimental data for calibrating the spatial and temporal explicit models are scarce.
Jakobsen and collegues \cite{jakobsen1992external} presented measurements of hyphal length densities of three AM fungi: \textit{Scutellospora calospora} (Nicol.\& Gerd.) Walker \& Sanders; \textit{Glomus} sp. associated with clover (\textit{Trifolium subterraneum} L.);these data appeared suitable for comparison with modelled hyphal length densities.

The model in  \cite{schnepf2008growth} describes, by means of a system of Partial Differential Equations (PDE), the development and distribution of the fungal mycelium in soil in terms of the creation and death of hyphae, tip-tip and tip-hypha anastomosis, and the nature of the root-fungus interface. It is calibrated and corroborated using published experimental data for hyphal length densities at different distances away from root surfaces. A good agreement between measured and simulated values was found for the three fungal species with different morphologies associated with \textit{Trifolium subterraneum} L. The model and findings are expected to
contribute to the quantification of the role of AM fungi in plant mineral nutrition and the
interpretation of different foraging strategies among fungal species.

\input{model}

%% file: model.tex
\subsection{Surface CWC Model}

In this Section we describe how to model the growth of arbuscular mycorrhyzal fungi using the surface spatial CWC.
We model the  growth of AM fungal hyphae in a soil environment partitioned into $13$ different layers (spatial compartments with label $soil$) to account for the distance in centimetres between the plant root and the fungal hyphae where the soil layer at the interface with the plant root is at position~$1.1$. We describe the mycelium by two atoms:
the hyphae (atom $Hyp$) related to the length densities (number of hyphae in a given compartment) and the hyphal tips (atom $Tip$). The plant root (atom $Root$) is contained in the $soil$ compartment at position $1.1$.

The tips and hyphae at the root-fungus interface proliferate according to the following spatial events:

$$\{ \mathtt{1.1}\} \triangleright soil : Root \srewrites{\tilde{a}} soil : Root \conc Hyp $$
$$\{ \mathtt{1.1}\} \triangleright soil : Root \srewrites{a} soil : Root \conc Tip $$

\noindent where $\tilde{a}$ and $a$ is the root proliferation factor for the hyphae and tips respectively.

Hyphal tips are important, because growth occurs due to the elongation of the region just behind the tips. Therefore, the spatial movement event describing the hyphal segment created during a tip shift to a nearby compartment is:\\

$$
[*] \triangleleft  \{\mathtt{E}, \mathtt{W}\} \triangleright soil, soil: Tip, \emptyseq \srewrites{v} soil, soil: Hyp , Tip
$$

\noindent where $v$ is the rate of tip movement. The hyphal length is related to tips movement, i.e. an hyphal trail is left behind as tips move through the compartments. We consider hyphal death to be linearly
proportional to the hyphal density, so that
the rule describing this spatial event is:\\

$$ [*] \triangleright soil : Hyp \srewrites{d_H} soil : \emptyseq $$

\noindent where $d_H$ is the rate of hyphal death.

Mycorrhizal fungi are known to branch mainly apically where one tip splits into two. In the simplest case, branching and
tip death are linearly proportional to the existing tips in that location modelled with the following spatial events:

$$ [*] \triangleright soil : Tip \srewrites{b_T} soil : 2*Tip $$
$$ [*] \triangleright soil : Tip \srewrites{d_T} soil : \emptyseq $$

\noindent where $b_T$ is the tip branching rate and $d_T$ is the tip death rate.

Alternatively, if we assume that branching decreases with increasing tip density and ceases at a given maximal tip density, we employ the spatial event:

$$ [*] \triangleright soil : 2* Tip \srewrites{c_T} soil : \emptyseq $$

\noindent where $c_T=\frac{b_T}{T_{max}}$. From a
biological point of view, this behaviour take into account the volume saturation when the tip density achieves the maximal number of tips $T_{max}$.

The fusion of two hyphal tips or a tip with a hypha
can create interconnected networks by means of
anastomosis:

$$ [*] \triangleright soil : 2* Tip  \srewrites{a_1} soil : Tip $$
$$ [*] \triangleright soil : Tip \conc Hyp \srewrites{a_2} soil : Tip $$

\noindent where $a_1$ and $a_2$ are the tip-tip and tip-hypha anastomosis rate constants, respectively.

The initial state of the system is given by the following grid cell definition:

$$\{ \mathtt{1.1}\},\;soil \boxplus  Root \conc T_0 * Tip \conc  H_0 * Hyp$$
$$\mathtt{rect}[\mathtt{1.2},\mathtt{1.13}],\;soil \boxplus  \emptyseq$$

\noindent where  $T_0$ and $H_0$ are the initial number of tips and hyphae respectively at the interface with the plant root.

\subsection{Results}

\begin{figure}
\center
\includegraphics[angle=-90,width=.49\textwidth]{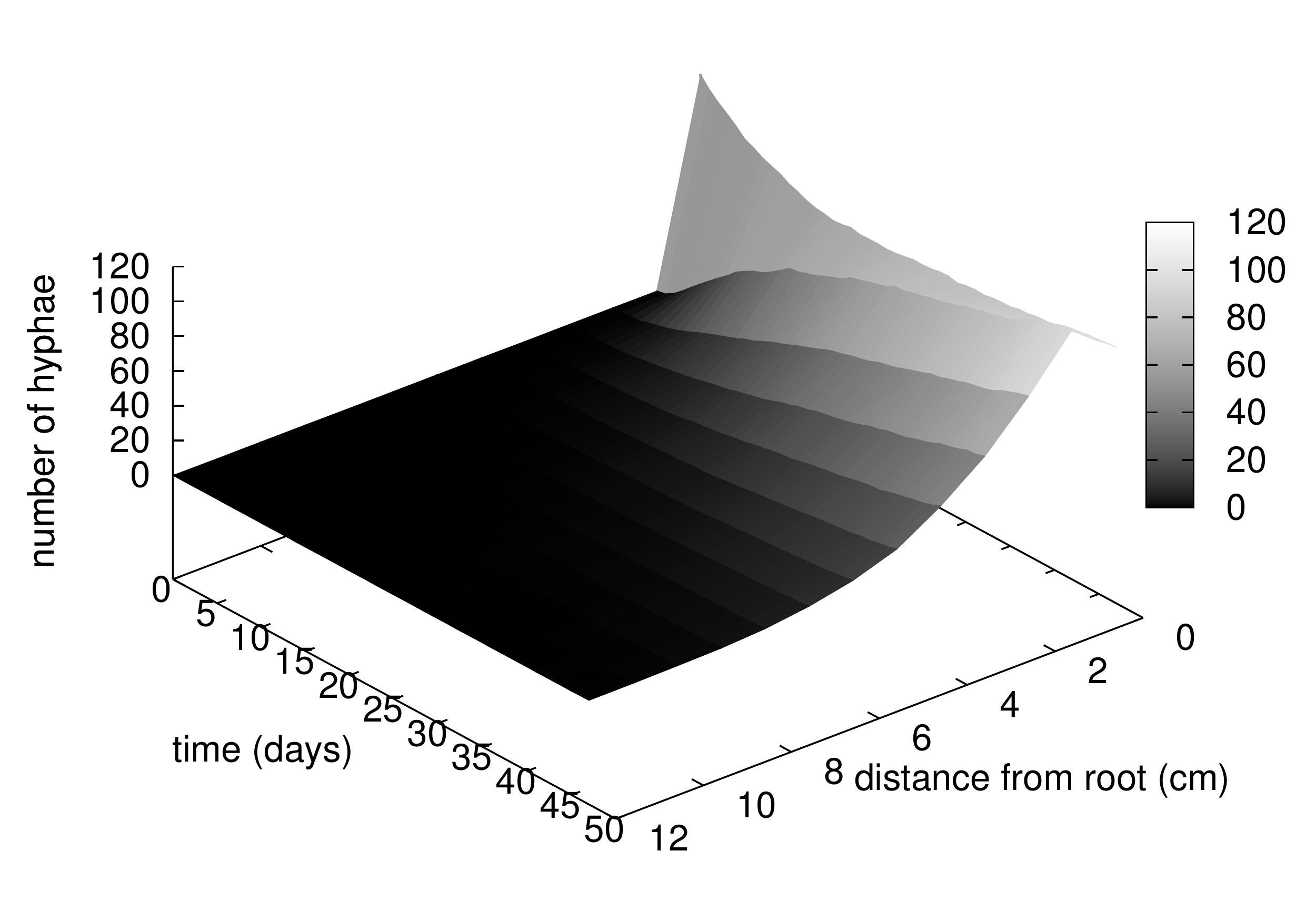}
\includegraphics[angle=-90,width=.49\textwidth]{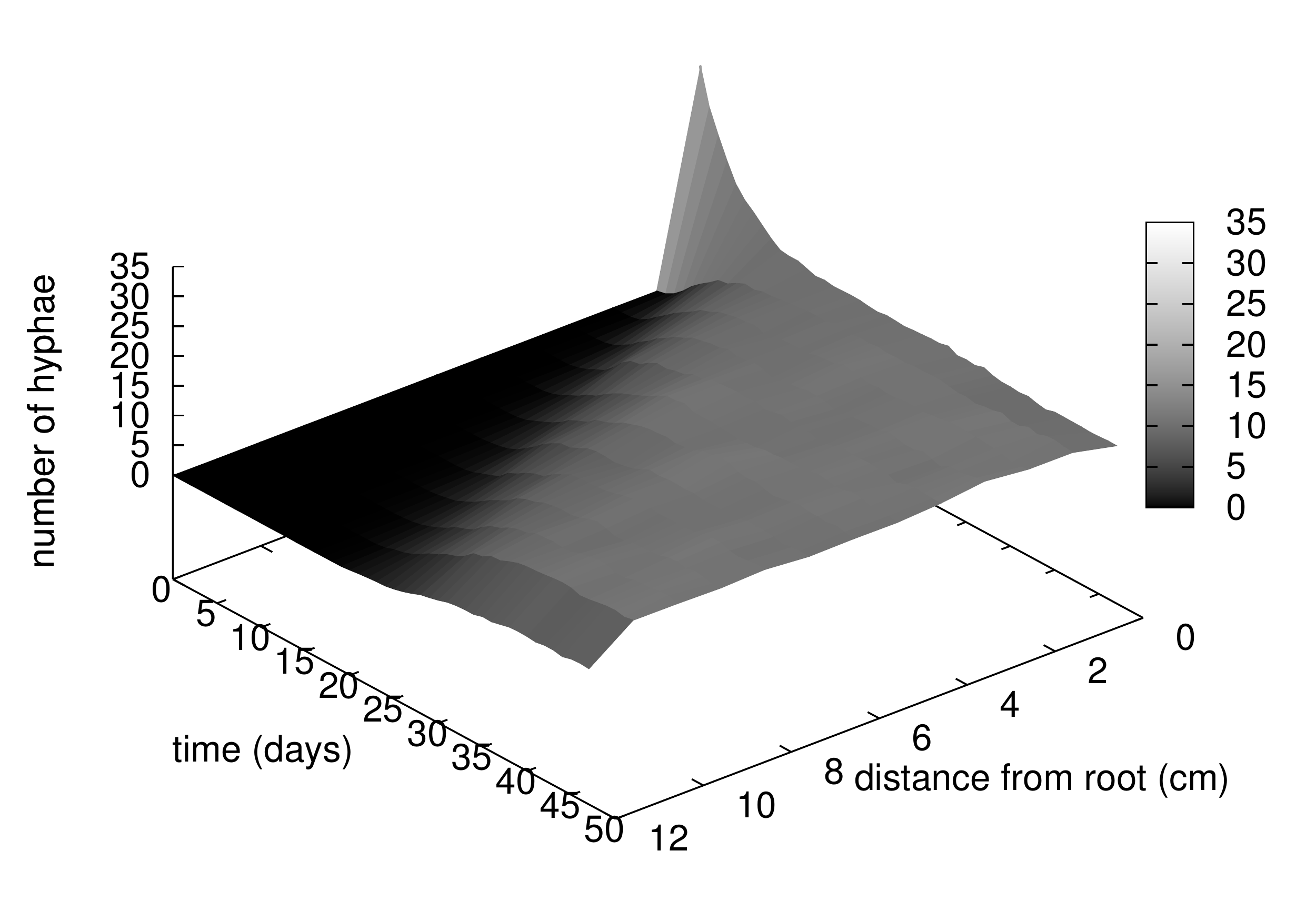}
\caption{Mean values of $60$ stochastic simulations of hyphal growth (atoms $Hyp$) results for S. calospora and Glomus sp. fungi.}\label{res_fungi}
\end{figure}

We run $60$ simulations on the model for the fungal species \textit{Scutellospora
calospora} and \textit{Glomus} sp. Figure \ref{res_fungi} show the mean values of hyphae (atoms $Hyp$) of the resulting stochastic simulations in function of the elapsed time in days and of the distance from the root surface. The rate parameters of the model are taken from~\cite{schnepf2008growth}.

The results for \textit{S. calospora} are in accordance with the linear PDE model of~\cite{schnepf2008growth}  which is characterized by linear branching with a relatively small net branching rate and both kinds of anastomosis are negligible when compared with the other species. This model imply that the fungus is mainly growing and allocating resources for
getting a wider catchment area rather than local expoloitation of mineral resources via hyphal branching.

The model for \textit{Glomus} sp. considers the effect of nonlinear
branching due to the competition between tips for space. The results obtained for \textit{Glomus} sp. are in accordance with the non--linear PDE model of~\cite{schnepf2008growth} which imply that local exploitation for resources via hyphal branching is important for
this fungus as long as the hyphal tip density is small. Reaching near the maximum tip density, branching decreases.
Symbioses between a given host plant and different AM fungi have been shown to differ functionally~\cite{ravnskov1995functional}.

%% file: conclu.tex
For the well-mixed chemical systems (even divided into nested
compartments) often found in cellular biology, interaction and
distribution analysis are sufficient to study the system's
behaviour. However, there are many other situations, like in cell
growth and developmental biology where dynamic spatial
arrangements of cells determines fundamental functionalities,
where a spatial analysis becomes essential. Thus, a realistic
modelling of biological processes requires space to be taken into account~\cite{K06}.

This has brought to the extension of many formalisms developed for the analysis of biological systems with (even continuous) spatial features.

In~\cite{CG10}, Cardelli and Gardner develop a calculus of processes located in a three-dimensional geometric space. The calculus, introduces a single new geometric construct, called {\it frame shift}, which applies a three-dimensional space transformation to an evolving process. In such a work, standard notions of process equivalence give rise to geometric invariants.

In~\cite{BMMPT11}, a variant of P-systems embodying concepts of space and position inside a membrane is presented. The objects inside a membrane are associated with a specific position. Rules can alter the position of the objects. The authors also define {\it exclusive} objects (only one exclusive object can be contained inside a membrane). In~\cite{BMMP09}, an spatial extension of CLS is given in a 2D/3D space. The spatial terms of the calculus may move autonomously
during the passage of time, and may interact when the constraints on their positions are satisfied. The authors consider a {\it hard-sphere} based notion of space: two objects, represented as spheres, cannot occupy the same space, thus conflicts may arise by moving objects. Such conflicts are resolved by specific algorithms considering the forces involved and appropriate pushing among the objects.

BioShape~\cite{BCCMT10} is a spatial, particle-based and multi-scale 3D simulator. It treats biological
entities of different size as geometric 3D {\it shapes}. A shape is either basic
(polyhedron, sphere, cone or cylinder) or composed
(aggregation of shapes glued on common surfaces of contact). Every element
involved in the simulation is a 3D process and has associated its physical motion law.

Adding too many features to the model (e.g., coordinates, position, extension, motion direction and speed, rotation, collision and overlap detection, communication range, etc.) could heavily rise the complexity of the analysis. To overcome this risk, a detailed study of the possible subsets of these features, chosen to meet the requirements of particular classes of biological phenomena, might be considered.

In this paper we pursued this direction by extending CWC with a surface language providing a framework for incorporating basic spatial features (namely, coordinates, position and movement). In future work we plan to extend the surface language to deal with three dimensional spaces and to investigate the possibility to incorporate other spatial features to the CWC simulation framework.

Notably, the framework presented in this paper could also be applied to other calculi which are able to express compartmentalisation (see, e.g., BioAmbients~\cite{RPSCS04}, Brane Calculi~\cite{Car04}, Beta-Binders~\cite{DPPQ06}, etc.). 

%% file: implementation.tex
This Section presents a software module implementing the translation of a surface language model into the corresponding standard CWC model that can be executed by the CWC simulator (cf.\ Sec.~\ref{sec:simulator}). The module is written in Java by means of the ANTLR parser generator \cite{antlr:web}. The input syntax of the software is defined as following.

\smallskip \noindent
{\it Patterns, Terms and Open Terms:} pattern, terms and open terms follow the syntax of CWC. In the definition of a compartment, its label is written in braces, as the first element in the round brackets. The symbol $\into$ is translated into $|$\lb{, and the empty sequence $\emptyseq$ into \texttt{$\backslash$e}}. If a pattern, term or open term is repeated several times, we write the number of repetitions before it.

\smallskip \noindent
{\it Grid Coordinates:} the row and the column of a grid coordinate are divided by a comma. All the constructions of the surface language are implemented. The components of a set of coordinates are divided by a blank space.

\smallskip \noindent
{\it Directions:} for the directions we use the same keywords of the surface language, plus the special identifiers \texttt{+}, \texttt{x}, \texttt{*} to identify all the orthogonal directions (\texttt{N},\texttt{S},\texttt{W},\texttt{E}), all the diagonal directions (\texttt{NW},\texttt{SW},\texttt{NE},\texttt{SE}) and all the directions, respectively.

\smallskip \noindent
{\it Model Name:} the name of the model is defined following the syntax\\
\centerline{\texttt{model} $string$ \texttt{;}}\\
where $string$ is the name of the model.

\smallskip \noindent
{\it Grid Dimensions:} the dimensions of the grid are expressed with the syntax\\
\centerline{\texttt{grid} $r$ \textbf{,} $c$ \texttt{;}}\\
where $r$ and $c$ are the number of rows and columns of the grid, respectively.

\smallskip \noindent
{\it Grid Cell Construction:} the notation of a grid cell $\Theta, \ell_s  \boxplus \; \ov{t} $ is translated into the code line\\
\centerline{\texttt{cell} \texttt{<} $\Theta$ \texttt{>} \texttt{\{} $\ell_s$ \texttt{\}} $\ov{t}$ \texttt{;}}\\
\lb{the module writes as many CWC compartments as the number of coordinates in $\Theta$: each of these copies has the same label $\ell_s$ and the same content $\ov{t}$, but a different coordinate in the wrap.}

\smallskip \noindent
{\it Non Spatial Events:} the notation of a non spatial event $\ell:  \ov\LeftPat \srewrites{k} \ov\RightPat $ is translated into the code line\\
\centerline{\texttt{nse} \texttt{\{} $\ell$ \texttt{\}} $\ov\LeftPat$ \textbf{[} $k$ \textbf{]} $\ov\RightPat$ \texttt{;}}\\
\lb{the module translates this line in a unique CWC rule.}

\smallskip \noindent
{\it Spatial Events:} the notation of a spatial event $\Theta \triangleright \ell_s : \overline{\LeftPat} \srewrites{k}  \ell_s': \ov\RightPat$ is translated into the code line\\
\centerline{\texttt{se} \texttt{<} $\Theta$ \texttt{>} \texttt{\{} $\ell_s$ \texttt{\}} $\ov\LeftPat$ \textbf{[} $k$ \textbf{]} \texttt{\{} $\ell_s'$ \texttt{\}} $\ov\RightPat$ \texttt{;}}\\
As shortcuts, the absence of \texttt{<} $\Theta$ \texttt{>} indicates the whole grid, and the absence of \texttt{\{} $\ell_s'$ \texttt{\}} indicates that the label of the spatial compartment does not change. \lb{The module writes a CWC rule for each coordinate in~$\Theta$: a rule differs from the others only in the coordinate written in its wrap.}

\smallskip \noindent
{\it Spatial Movement Events:} the notation of a spatial movement event $\Theta \triangleleft \Delta \triangleright \ell_{s_1}, \ell_{s_2}: \ov{p_1} ,\ov{p_2} \srewrites{k} \ell_{s_1}', \ell_{s_2}': \ov{o_1} ,\ov{o_2}$ is translated into the code line\\
\centerline{\texttt{sme} \texttt{<} $\Theta$ \texttt{>} \texttt{[} $\Delta$ \texttt{]} \texttt{\{} $\ell_{s_1}$ \texttt{\}} $\ov{p_1}$ \texttt{\{} $\ell_{s_2}$ \texttt{\}} $\ov{p_2}$ \textbf{[} $k$ \textbf{]} \texttt{\{} $\ell_{s_1}'$ \texttt{\}} $\ov{o_1}$ \texttt{\{} $\ell_{s_2}'$ \texttt{\}} $\ov{o_2}$ \texttt{;}}\\
As shortcuts, the absence of \texttt{<} $\Theta$ \texttt{>} indicates the whole grid, and the absence of \texttt{\{} $\ell_{s_1}'$ \texttt{\}} or \texttt{\{} $\ell_{s_2}'$ \texttt{\}} indicates that the label of the spatial compartment does not change. In case of absence of \texttt{\{} $\ell_{s_2}'$ \texttt{\}}, an underscore is used to separate $\ov{o_1}$ and $\ov{o_2}$. \lb{For each coordinate in $\Theta$, the module writes as many CWC rules as the number of directions in $\Delta$; in case of a coordinate on the edge of the grid, the module writes a CWC rule for a direction only if this one identifies an adjacent spatial compartment on the grid. The number of CWC rules is therefore less or equal to $|\Theta| \times |\Delta|$.}

\smallskip \noindent
{\it Monitors:} a monitor permits to expose what pattern we need to monitor: at the end of simulation, all the states of this pattern are written in a log file. The syntax to design a monitor is the following:\\
\centerline{\texttt{monitor} $string$ \texttt{<} $\Theta$ \texttt{>} \texttt{\{} $\ell_s$ \texttt{\}} $\ov\LeftPat$ \texttt{;}}\\
where $string$ is a string describing the monitor and $\ov\LeftPat$ is the pattern, contained into a spatial compartment labelled by $\ell_s$ and the coordinates $\Theta$, to monitor. As shortcut, the absence of \texttt{<} $\Theta$ \texttt{>} indicates the average of the monitors in the whole grid, and the absence of \texttt{\{} $\ell_s$ \texttt{\}} indicates to write a monitor for each label defined in the model. \lb{The module writes a monitor for each coordinate in $\Theta$; in case of the absence of~\texttt{\{}$\ell_s$\texttt{\}}, the module writes a monitor for each combination of coordinates in $\Theta$ and spatial labels defined in the model.}

The construction of a model follows the order used to describe the translator syntax: first the model name and the grid dimensions, then the rules of the model. After the rules, we define the grid cell, and finally the monitors.

\begin{lstlisting}[float=t,label=input_sl1,caption=Input file for the Surface Language parser to model the \textit{S. calospora} fungus growth,frame=single]
model "AM Fungi Growth Model S. Calospora";

grid 1,13;

//  Tips and hyphae proliferation at the root-fungus interface
se <0,0> {grid} Root [2.52] Root Tip;
se <0,0> {grid} Root [3.5] Root Hyp;

// Tips branching and death
se {grid} Tip [0.02] 2 Tip;
se {grid} Tip [0.0052] \e;

// Hyphae death
se {grid} Hyp [0.18] \e;

// Hyphal creation during a tip shift to a nearby compartment
sme [E W] {grid} Tip {grid} \e [0.125] Hyp _ Tip;

//  Initial state
cell <0,0> {grid} Root 97 Tip 115 Hyp;	
cell <rect[0,1 0,12] >  {grid} \e;

monitor "Hyp" <rect[0,0 0,12] > {grid} Hyp;
\end{lstlisting}

\begin{lstlisting}[float=t,label=input_sl2,caption=Input file for the Surface Language parser to model the \textit{Glomus} sp. fungus growth,frame=single]
model "AM Fungi Growth Model Glomus sp.";

grid 1,13;

//  Tips and hyphae proliferation at the root-fungus interface
se <0,0> {grid} Root [2.24] Root Tip;
se <0,0> {grid} Root [1.04] Root Hyp;

// Tips branching and death
se {grid} Tip [1.91] 2 Tip;
se {grid} Tip [0.15] \e;

// Tips saturation
se {grid} 2 Tip [0.15] \e;

// Hyphae death
se {grid} Hyp [0.28] \e;

// Hyphal creation during a tip shift to a nearby compartment
sme [E W] {grid} Tip {grid} \e [0.065] Hyp _ Tip;

//  Initial state
cell <0,0> {grid} Root 84 Tip 35 Hyp;	
cell <rect[0,1 0,12] >  {grid} \e;

monitor "Hyp" <rect[0,0 0,12] > {grid} Hyp;
\end{lstlisting}

Listings \ref{input_sl1} and \ref{input_sl2} show the input file  for the CWC Surface Language software to model the {\it S. calospora} and  {\it Glomus} sp. fungi growth.